\documentclass[aps,prl,twocolumn,floatfix,superscriptaddress]{revtex4-1}
\usepackage{newtxtext,newtxmath}
\usepackage{mathrsfs}
\usepackage[export]{adjustbox}
\usepackage[usenames,dvipsnames]{color}
\usepackage[colorlinks=true,citecolor=blue,breaklinks=true,linkcolor=red,linktocpage=true,urlcolor=blue,pagebackref=false]{hyperref}

\usepackage{graphicx}

\newcommand{\ket}[1]{|#1\rangle}
\newcommand{\bra}[1]{\langle#1|}

\begin{document}

\title{Single atom laser in normal-superconductor quantum dots}
\author{Gianluca Rastelli}
\affiliation{Fachbereich Physik, Universit{\"a}t Konstanz, D-78457 Konstanz, Germany}
\affiliation{Zukunftskolleg, Universit{\"a}t Konstanz, D-78457 Konstanz, Germany}
\author{Michele Governale}
 \affiliation{School of Chemical and Physical Sciences and MacDiarmid Institute for Advanced Materials and Nanotechnology,
Victoria University of Wellington, PO Box 600, Wellington 6140, New Zealand}
\date{\today}

\begin{abstract}
We study a single-level quantum dot strongly coupled to a superconducting lead and tunnel-coupled to a normal 
electrode which  can exchange energy with a single-mode resonator.
We show that a such system can sustain lasing characterized by 
a sub-Poissonian Fock-state distribution of the resonator.
The lasing regime is clearly identifiable in the subgap transport regime: in the resonant case, the current is 
pinned to the maximum value achievable in this hybrid nanostructure.
\end{abstract}

\maketitle

%
%
%
{\sl Introduction.}
%
%
Hybrid nanoscale systems combine elemental components of different natures.
They provide a way to explore novel mechanisms of coherent energy exchange.  
Experimental systems that are commonly studied comprise 
 quantum dots coupled to localized harmonic oscillators such as 
microwave photon cavities\cite{Viennot:2015ir,Stockklauser:2015,Bruhat:2016br,Mi:2016ex,Stockklauser:2017bqa,Li:2018jr}
or 
mechanical resonators \cite{Benyamini:2014eb,Okazaki:2016bh,Deng:2016iw,Urgell:2019}.
They display a variety of different dynamic behaviors related to  quantum transport in the single-electron regime.
In particular, these systems can realize single atom 
lasers \cite{Walther:2006da,Childress:2004kt,Jin:2011gs,Kulkarni:2014}
which have been experimentally demonstrated 
in double quantum  dots coupled to microwave cavities \cite{Liu:2015bha,Liu:2017}. 
Beside photon cavities, nanomechanical  systems can also sustain {\sl phonon} lasing, 
as, for instance, in a superconducting single-electron-transistor setup \cite{Rodrigues:2007}. 
Single atoms lasers  exhibit unique features compared to conventional lasers, such as,  
multistability \cite{Walther:2006da,Rodrigues:2007fv,Mantovani:2019,Nation:2013db,Lambert:2015}. 
Furthermore, such hybrid nanoscale systems are ideal platforms to explore correlations between charge transport and nonequilibrium 
photon states \cite{Zhang:2012ie,Cottet:2012,Souquet:2014gm,Stadler:2016,Stadler:2017,Arrachea:2014cm}.

%
%
%
\begin{figure}[t!]
\includegraphics[scale=0.115]{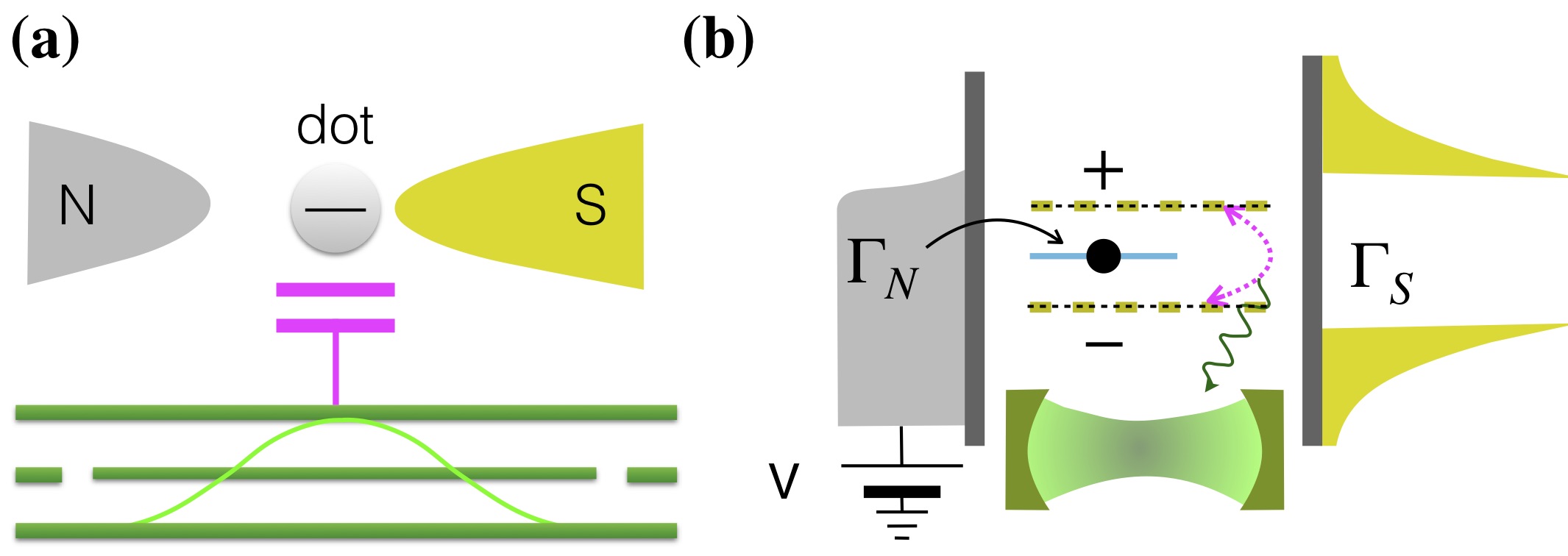}\\[3mm] 
\includegraphics[scale=0.115]{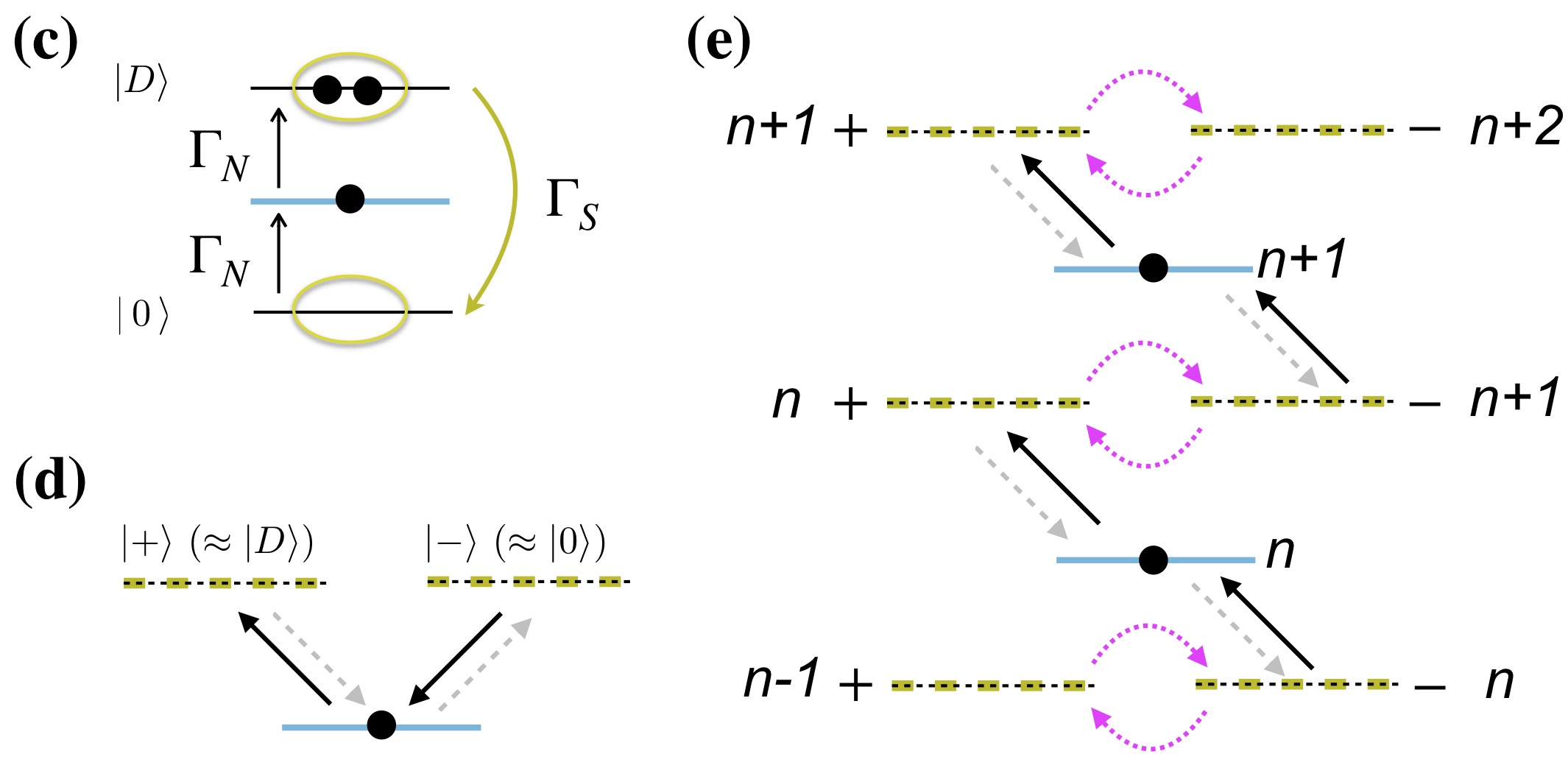}\\
%
%
\caption{
{\bf (a)} 
Device setup: a quantum dot between a superconductor  (S) and a normal metal (N) 
is capacitively coupled to a microwave cavity. 
{\bf (b)} 
In the the rotating wave approximation, the states $\ket{+}$ and $\ket{-}$ indicated by
the (yellow) dashed lines are dynamically coupled  to the resonator. 
The (blue) solid line denotes the singly-occupied states $\ket{\sigma}$.
$\Gamma_N/\hbar$ is the tunnelling rate from $N$  and $\Gamma_S$ is the effective pairing potential in the quantum dot.
{\bf (c)} 
In the high-bias voltage limit,  only the empty state $\ket{0}$ 
can transition to the singly-occupied state which in turn can transition to the doubly-occupied state $\ket{D}$.
{\bf (d)}  
At positive and large detuning $\delta$, the states $\ket{+} \approx \ket{D} - \frac{\Gamma_S}{2 \delta}\ket{0}$ and $\ket{-}\approx\ket{0} + \frac{\Gamma_S}{2\delta}\ket{D}$.
This breaks the symmetry in the transitions between the states $\ket{\pm}$  and  $\ket{\sigma}$.
{\bf (e)}  
Description of the pumping mechanism using the level-diagram of the quantum dot coupled to the resonator (see text).
}
\label{fig:cartoon}
\end{figure}
%
%
%

In this letter we address the question of whether single-atom lasing can be achieved in a quantum dot with a single orbital level, strongly coupled to a superconducting lead. 
In  Fig.~\ref{fig:cartoon}(a,b) we show a specific example with a microwave photon cavity, similar to the experimental set up of Ref.\cite{Bruhat:2016br}. 

Before discussing our results in detail, we wish to give an intuitive description of how lasing  occurs in this system.
In the limit of a large superconducting gap, Andreev bound states, $\ket{+}$ and $\ket{-}$, are formed in the dot.
These are coherent superpositions of the empty and doubly-occupied dot states, $\ket{0}$ and $\ket{D}$, respectively.
At high bias voltage, the normal lead behaves as a source of electrons, hence only the empty state $\ket{0}$ 
can transition by tunnelling to one of the singly-occupied states, $\ket{\sigma}$, which in turn can transition to the doubly-occupied state $\ket{D}$, Fig.~\ref{fig:cartoon}(c). 
At sufficiently large energy detuning, $\delta$, between the two states $\ket{D}$ and $\ket{0}$, 
the state $\ket{-}$  has a larger component of the state $\ket{0}$,  
whereas  $\ket{+}$ is mostly the state $\ket{D}$.
This implies that the transitions $\ket{-} \rightarrow \ket{\sigma} $ and $\ket{\sigma} \rightarrow \ket{+}$ are favoured over the opposite transitions, Fig.~\ref{fig:cartoon}(d). 
This symmetry breaking is ultimately the reason for lasing.
Additionally, the product states corresponding to different occupations of the resonator 
(Fock states) and the Andreev bound states, $\ket{\pm,n}\equiv \ket{\pm}\otimes \ket{n}$, 
are  mixed by the interaction between the charge on the dot and the resonator mode 
[curved dashed arrows in Fig.~\ref{fig:cartoon}(e)].
As explained above, for the right choice of system parameters (detuning $\delta$), tunnelling with the normal lead 
favours the chain of transitions  $\ket{-,n} \rightarrow \ket{\sigma,n} \rightarrow\ket{+,n}$, in which 
the state of the resonator is not modified.
The interaction with the resonator then brings $\ket{+,n}$ into $\ket{-,n+1}$ and the process can start again. This leads to an increasingly higher occupation of the resonator mode, Fig.~\ref{fig:cartoon}(e). 
Eventually, this energy-pumping mechanism is balanced by the intrinsic losses of the resonator 
and hence a steady state with a large but finite average occupation $\bar{n}$ is established.  
More precisely, we show that the resonator reaches a lasing state in the steady-state regime.
In particular, by tuning the orbital energy level of the dot $\varepsilon_0$, 
one can control the energy splitting $2\epsilon_A$ between the state $\ket{+}$ 
and $\ket{-}$.
Accordingly, one can achieve the resonant condition for the coupling between the dot's degree of freedom and the 
resonator mode when  $2\epsilon_A = \hbar \omega_0$,  with $ \omega_0$ the resonator's frequency. 

We employ the Linblad-equation formalism in the rotating wave approximation (RWA) and find, in the semiclassical limit, 
the threshold value for the coupling strength between the dot and the resonator beyond which lasing occurs, 
namely $\lambda > \lambda_c$. 
We supplement the semiclassical treatment by a quantum master-equation approach, in order to evaluate the effect of quantum fluctuations.
The master equation approach predicts a sub-Poissonian Fock distribution for the resonator.
Remarkably, we find that the resonator's lasing is clearly visible in the transport properties.
Close to the resonant regime, the current is  pinned 
to the maximum value that characterizes the system in the large-gap limit. 
This pinning occurs independently of the system parameters, provided  that  the resonator is in the lasing state.

%
%
%
{\sl Model and formalism.}
%
%
The Hamiltonian of the system can be written as 
$H = H_{\text{osc}}+H_\text{int} + H_{\text{dS}} + H_\text{tun}  + H_{\text{N}} .$
The quantum dot coupled to the superconductor in the large-gap limit  is described by the effective Hamiltonian \cite{Rozhkov:2000}
\begin{align}
\label{eq:HdS}
H{_\text{dS}} &= \varepsilon_0 \sum_{\sigma=\uparrow,\downarrow} n_\sigma +U n_{\uparrow}n_{\downarrow}
-\frac{\Gamma_{S}}{2} \left( d^\dagger_{\uparrow} d^\dagger_{\downarrow}  + d_{\downarrow}^{\phantom{\dagger}}  d_{\uparrow}^{\phantom{\dagger}}  \right) 
\, , 
\end{align}
where $d_{\sigma} (d_{\sigma}^\dagger)$ is the annihilation(creation) operator for an electron with spin
$\sigma=\uparrow,\downarrow$ and $n_{\sigma}=d^{\dagger}_{\sigma}d_{\sigma}$. 
The Hilbert space of the dot is spanned by  the states $|0\rangle$ (empty), $|\sigma\rangle=d_\sigma^\dagger|0\rangle$ 
(singly occupied with spin $\sigma$), and 
$|D\rangle= d_\uparrow^\dagger d_\downarrow^\dagger |0\rangle$ (doubly occupied). 
The detuning between the energy of the doubly occupied state and the empty state $\delta = 2\varepsilon_0 + U$ determines the proximity effect in the dot. 
The eigenstates of the effective Hamiltonian are the singly-occupied states  $\ket{\sigma}$ with eigenenergies $E_\sigma=\varepsilon_0$ 
and the states $\left| \pm \right>$  which are coherent superpositions of  $\left| D \right>$   and  $\left| 0 \right>$  and read 
\begin{align}
\left| + \right> & = \cos(\theta) \left| D \right>  -  \sin(\theta) \left| 0 \right>  \, , \\
\left| - \right> & =  \sin(\theta) \left| D \right>  +  \cos(\theta) \left| 0 \right>  \, , 
\end{align}
with energies 
$E_{\pm} = \delta/{2} \pm \epsilon_A $, 
and coefficients 
$\cos(\theta) = ({1}/{\sqrt{2}}) {[1 + \delta/(2\epsilon_A)}]^{1/2}$ and 
$\sin(\theta) = ({1}/{\sqrt{2}}) {[ 1 - \delta/(2\epsilon_A)]}^{1/2}$.
The energy  $2\epsilon_A=\sqrt{\delta^2+\Gamma_S^2}$ is the splitting between the $|+\rangle$ and the $|-\rangle$ states. 
The normal lead is described by 
$H_{\text{N}} =
\sum_{k \sigma} \left(  \varepsilon_{k} - \mu_N \right)  c_{k \sigma}^\dagger c_{k \sigma}^{\phantom{\dagger}}$, 
with the lead-electron operators $c_{ k \sigma}^{\phantom{\dagger}} $ and $c_{k \sigma}^\dagger$. 
We choose as energy reference, the chemical potential of the superconductor $\mu_S$ and set without loss of generality 
$\mu_S=0$. The chemical potential of the normale lead is $\mu_N =  eV$, where $e$ is the electron charge and $V$ the voltage difference between the normal lead and the the superconductor.
The tunnelling between the lead and the dot is modelled by the tunnelling Hamiltonian
$H_{\text{tun}}  =
V \sum_{k \sigma} c^{\dagger}_{k\sigma} d^{\phantom{\dagger}}_{\sigma}  \, + \, \mathrm{H.c.}
$
We assume that the density of states $\rho_N$ of the normal lead  is constant in the energy window relevant for transport 
(wide-band approximation) and  define the tunnel-coupling strength as $\Gamma_N= 2\pi \rho_N  |V|^2$.
The Hamiltonian of the resonator reads
$H_\text{osc}= \hbar\omega_0 \,  a^\dagger a$ . 
The interaction Hamiltonian between the dot's degrees of freedom and the oscillator is
\begin{equation}
H_{\text{int}}=
\lambda \left(a^\dagger+a\right) \left( \sum_\sigma d_{\sigma}^\dagger d_\sigma^{\phantom{\dagger}}  -1 \right) \, ,
\end{equation}
where $\lambda$ denotes the interaction strength. 
We have defined the equilibrium position of the oscillator when the dot is singly occupied.
As shown in the Supplemental Material  \cite{sup_mat},
we use a polaronic transformation to decompose the charge-resonator interaction into a transverse 
and a longitudinal part with respect to the Andreev states.
The longitudinal interaction appears only in the tunnelling term \cite{Braig:2003cb} and leads the resonator 
to a nonequilibrium state with an average energy larger than in the thermal state \cite{Koch:2006vq}.
The inelastic tunneling processes associated with the longitudinal part scales as $\sim \lambda/(\hbar\omega_0)$ 
and one can safely neglect them for $\lambda \ll\hbar \omega_0$.
For the transverse interaction, we  introduce the the rotating-wave approximation (RWA), 
which is valid when $\hbar\omega_0\approx \left( E_{+} - E_{-}\right)  = 2 \epsilon_A$.
In the RWA the interaction Hamiltonian reads
\begin{equation}
\label{eq:HRWA}
H_{\text{int}}^{\text{RWA}}=
 \lambda  \sin\left(2\theta\right) 
 \Bigl[ a^\dagger \ket{-}\bra{+}+a \ket{+}\bra{-}\Bigr] \, .
\end{equation}
We define the Hamiltonian of the system $H_s=H_{\text{dS}}+H_{\text{osc}}+H_{\text{int}}^{\text{RWA}}$.
The reduced  density matrix of the quantum dot coupled to the oscillator, $\rho_{\text{s}}$,  is obtained by tracing out the normal lead and it obeys the following Lindblad equation
\begin{align}
\hbar\frac{\partial \hat{\rho}_{\text{s}} }{\partial t}
& =
-i\left[  H_{\text{s}} , \rho_{\text{s}} \right] - \frac{\Gamma_N}{2} 
\sum_{\sigma} \left[  \left\{ d_{\sigma}^{\phantom{\dagger}} d^{\dagger}_{\sigma} ,  \rho_{\text{s}}   \right\}
-2  d^{\dagger}_{\sigma}  \rho_{\text{s}}   d_{\sigma}^{\phantom{\dagger}}  \right]
\nonumber \\
& 
- \frac{\kappa}{2} 
 \left[  \left\{ a^{\dagger} a^{\phantom{\dagger}}  ,  \rho_{\text{s}}   \right\}
-2  a^{\phantom{\dagger}}   \rho_{\text{s}}   a^{\dagger}\right]
\label{eq:Lindlblad}
\end{align}
with $\left\{  ,  \right\}$ the anticommutator.
The second term of the  Lindblad  Eq.~(\ref{eq:Lindlblad}) describes the single electron tunnelling with the normal lead and 
is valid in the high-voltage regime with $eV>0$, that is when   
$eV$ is much larger than all the other energy scales of the system (except the superconducting gap) \cite{Mantovani:2019}.
The third term of Eq.~(\ref{eq:Lindlblad}) describes the relaxation of the oscillator towards its ground state quantified by a rate $\kappa/\hbar$
(namely the intrinsic finite losses of the resonator).

%
%
%
%
{\sl Semiclassical approximation.}
%
Assuming the resonator to be in a lasing state, one can use the mean-field or semiclassical approximation in which 
we replace the bosonic operator $a$ with its expectation value $a \approx \left< {a} \right> = \alpha$.
The quantity ${|\alpha|}^2$ represents, in the semiclassical language, the average  number of bosons, i.e. 
${|\alpha|}^2 \equiv A^2 \approx \left< a^{\dagger} a^{\phantom{\dagger}} \right> = \bar{n}$.
In terms of the Fock states, the condition of validity for the semiclassical approximation reads $\bar{n} \gg \delta n$ with 
$\delta n =[\langle \, (a^{\dagger} a^{\phantom{\dagger}})^2 \,  \rangle - \bar{n}^2]^{1/2}$.
Within the semiclassical approach, one can write a closed system of equations for the resonator amplitude $\alpha$ and the matrix elements 
of the reduced density matrix of the dot  \cite{sup_mat}.
We focus on the resonant regime $\hbar\omega_0=2\epsilon_A$.
The equation for the oscillator's energy dynamics  is given by
\begin{equation}
\label{eq:A_dynamics}
\frac{\partial A^2 }{\partial t}
\! = \! 
\frac{1}{\hbar}
\left[ - \kappa -  \gamma_{\textrm{eff}}\left( A^2 \right) \right] A^2 \, ,
\end{equation}
with 
\begin{equation}
\label{eq:gamma_eff}
\gamma_{\textrm{eff}}\left (A^2 \right)
=
 -  \frac{2 \lambda^2}{\Gamma_N} \,
\frac{\sin^2(2\theta) \cos(2\theta)}{1 +  \frac{4 \, \lambda^2 \sin^2(2\theta)}{\Gamma_N^2}  \, A^2 } \, .
\end{equation}
The Eq.~(\ref{eq:A_dynamics}) and Eq.~(\ref{eq:gamma_eff})  are the mean-field semiclassical lasing equations \cite{scully1997quantum}.
They describe the laser as a classical oscillator with a simple linear friction (with damping coefficient $\kappa$) and 
and a {\sl nonlinear} gain which arises from the interaction with the two quantum levels.
A  nontrivial steady-state solution $\bar{A} \neq 0$ exists for the coupling strength larger than a critical value $\lambda > \lambda_c$
which is known as the lasing threshold:
\begin{equation}
\lambda_c
=
\sqrt{\frac{\kappa \Gamma_N}{2}} 
\frac{ {\left[ \delta^2 + \Gamma_S^2 \right]}^{3/4} } { \Gamma_S \sqrt{\delta} }   
\, , \,\, \mbox{for} \,\, \delta>0.
\end{equation}
Notice that lasing can be achieved only for $\delta>0$ by choice of the applied voltage. 
If we invert the applied voltage, lasing occurs for $\delta<0$.
The solution for the amplitude  in the semiclassical approach, to be compared with the average oscillator occupation 
$(A^2={|\alpha|}^2  \approx \bar{n})$, is 
\begin{equation}
\bar{A}^2 =  \left(\frac{\Gamma_N}{2 \kappa}\right)  \frac{\delta}{\sqrt{\delta^2 + \Gamma_S^2}} \left[ 1 -  {\left( \frac{\lambda_c}{\lambda}\right)}^2 \right] \, .
\end{equation}
An adiabatic approximation which is valid for ultra-low damping  \cite{scully1997quantum} allows us to calculate the steady state population of the Fock states and 
the Fano factor of the resonator  $F = [\langle \, {(   a^{\dagger} a^{\phantom{\dagger}}  )}^2 \,  \rangle - \bar{n}^2] / \bar{n} $. 
For the present case, the adiabatic approximation  yields $F \simeq 1$ on resonance (more details in Sup. Mat. \cite{sup_mat}).

%
%
%
%
\begin{figure}[t!]
\begin{flushleft}
\includegraphics[scale=0.35,angle=270]{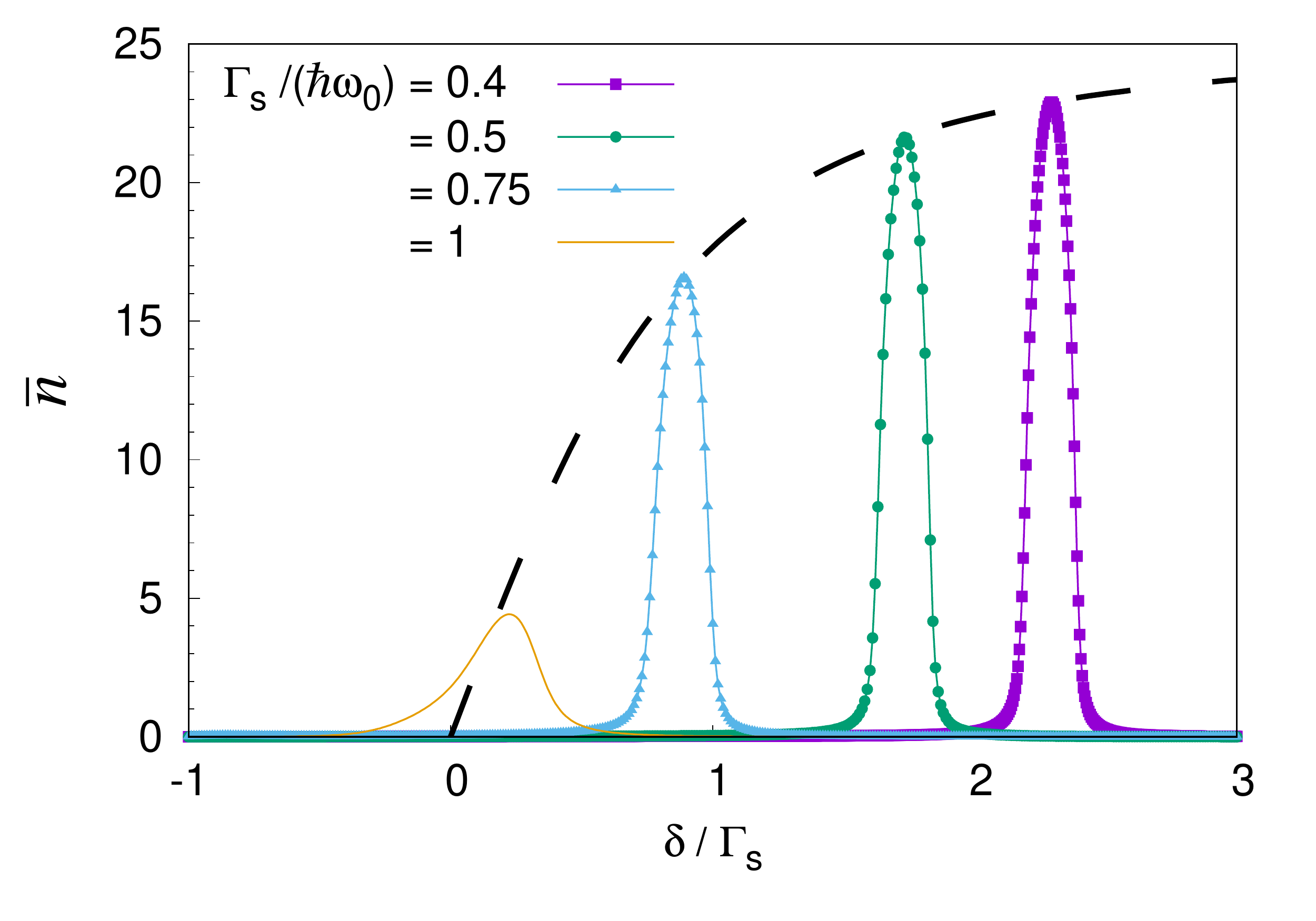}
\end{flushleft}
\caption{
Average occupation of the oscillator mode obtained by solution of the master equation.
$\bar{n}$ is plotted as  a function of the detuning $\delta=2\varepsilon+U$ in units of $\Gamma_S$ 
for different ratios of $\Gamma_S/(\hbar\omega_0)$.
The dashed line corresponds to the semiclassical approximation (in this case the system is always on resonance). 
Parameters: $\lambda=0.01\, \hbar\omega_0$ and $\kappa/\Gamma_N=0.02$.  
}
\label{fig:noschdiss}
\end{figure}
%
%
%

%
%
%
%
{\sl Quantum master equation in RWA} \cite{comparison}.
%
In order to assess the relevance of quantum effects for single-atom lasing in this system, we employ the quantum master equation approach. 
We now use the basis $\ket{s,n}=\ket{s}\otimes\ket{n}$ where  \linebreak $s\in\{\uparrow,\downarrow,+,-\}$ labels the eigenstates of the  
Hamiltonian in Eq.~(\ref{eq:HdS}).
Within the rotating-wave approximation the Hamiltonian 
$H_s$ can be diagonalised. 
The states $\ket{\sigma,n}$ are eigenstates of $H_s$ with eigenergies $E_{\sigma,n}=\varepsilon_0+\hbar\omega_0 n$. 
In presence of the interaction with the oscillator, the states $\left| \pm \right>$ are hybridized with the Fock states, yielding the following eigenstates
\begin{subequations}
\label{eq:states}
\begin{align}
\ket{{rw}{+},n}&= \sin(\varphi_n) \ket{+,n} - \cos(\varphi_n) \ket{-,n+1}  \, , \\
\ket{{rw}{-},n}&=\cos(\varphi_n) \ket{+,n} +  \sin(\varphi_n) \ket{-,n+1} \, , 
\end{align}
\end{subequations}
with $\sin(\varphi_n)= (1/{\sqrt{2}}) {[1- \Delta/( 2 W_n)]}^{1/2}$ and 
$\cos(\varphi_n)=(1/{\sqrt{2}}) {[1+ \Delta/( 2 W_n)]}^{1/2}$ 
where $\Delta=\hbar\omega_0-2\epsilon_A$ is 
the detuning between the states $\ket{-,n+1}$ and $\ket{+,n}$ 
and $W_n={[ \left(\Delta/2\right)^2+\lambda^2 \sin^2(2\theta) (n+1) ]}^{1/2}$.
The eigenenergies corresponding to the eigenstates given in Eq.~(\ref{eq:states}) are
$E_{rw\pm,n}=E_{+} +\hbar\omega_0 n+\Delta/{2}\pm W_n$.

Performing a perturbation expansion in $H_{\text{tunn}}$, one can obtain from the Lindblad Eq.~(\ref{eq:Lindlblad})  
a rate-equation for the populations of the eigenstates of $H_s$. 
The same master equation can be obtained in the framework of a diagrammatic real-time technique \cite{Pala:2007,Governale:2008}.
Here we restrict ourselves to first-order in $\Gamma_N$. 
We relabel the eigenstates of $H_s$ as $\ket{\alpha,n}$, with $\alpha \in \{\uparrow, \downarrow, rw+,rw-\} $. 
The master equation for the occupation probabilities $P_{\alpha,n}=\text{Tr}_{\text{lead}}\Bigr[\rho \ket{\alpha,n}\bra{\alpha,n}\Bigl]$, can be written as 
\begin{align}
\label{eq:master}
\dot{P}_{\alpha,n}=\frac{1}{\hbar}\sum_{\alpha',n'} W_{(\alpha,n);(\alpha',n')} P_{\alpha',n'}, 
\end{align}
where $W_{(\alpha,n);(\alpha',n')}/\hbar$, for $(\alpha',n') \ne (\alpha,n)$, is the transition rate from state $\ket{\alpha',n'}$ to  $\ket{\alpha,n}$. 
The diagonal elements of the kernel $W_{(\alpha,n);(\alpha',n')}$ are defined by $W_{(\alpha,n);(\alpha,n)}=-\sum_{(\alpha',n')\ne (\alpha, n)} W_{(\alpha',n');(\alpha,n)}$.

Using Fermi's golden rule, the transition rates associated to the tunnelling events with the normal lead in the high-bias regime (uni-directional transport) are 
\begin{align}
\label{eq:rates-hb}
W^{N}_{(\alpha,n);(\alpha',n')}(\chi) =\Gamma_{N} \sum_{\sigma}\Bigl[ e^{-i \chi}|\langle \alpha,n | d^{\dagger}_{\sigma} | \alpha',n'\rangle|^2\Bigr] \, ,
\end{align}
 where we have introduced the counting field $\chi$ in the usual way \cite{Braggio:2011}.
We introduce the damping of the resonator mode by coupling the oscillator to a zero-temperature bosonic bath. 
The corresponding rates are $W^{D}_{(\alpha,n);(\alpha',n')} =\kappa |\langle \alpha,n | a | \alpha',n'\rangle|^2$. The rates in the master equation (\ref{eq:master}) are given by $W_{(\alpha,n);(\alpha',n')}=W^{N}_{(\alpha,n);(\alpha',n')}(\chi)+W^{D}_{(\alpha,n);(\alpha',n')}$. The counting field $\chi$ needs to be removed from the diagonal elements of the kernel. 
The stationary probabilities, $P_{\alpha,n}^{\text{stat}}$, are obtained solving Eq.~(\ref{eq:master}) for $\dot{P}_{\alpha,n}=0$ 
with the condition that $\sum_{\alpha,n}P_{\alpha,n}=1$ and are the kernel of the matrix $W_{(\alpha,n);(\alpha',n')}$. 
The stationary current in the normal lead can be written in terms of the stationary probabilities as   
$I=-i\frac{e}{\hbar}\sum_{\alpha,\alpha',n,n'}\left.\frac{\partial W_{(\alpha,n);(\alpha',n')} }{\partial{\chi}}\right|_{\chi\rightarrow 0} P_{\alpha',n'}^{\text{stat}}$.
To perform numerical computations, we 
introduce a maximum value $n_\text{max}$ for the highest Fock state number with $n_\text{max} \gg \bar{n}$ \cite{sup_mat}.

%
%
%
%
\begin{figure}[t!]
\includegraphics[scale=0.34,angle=270]{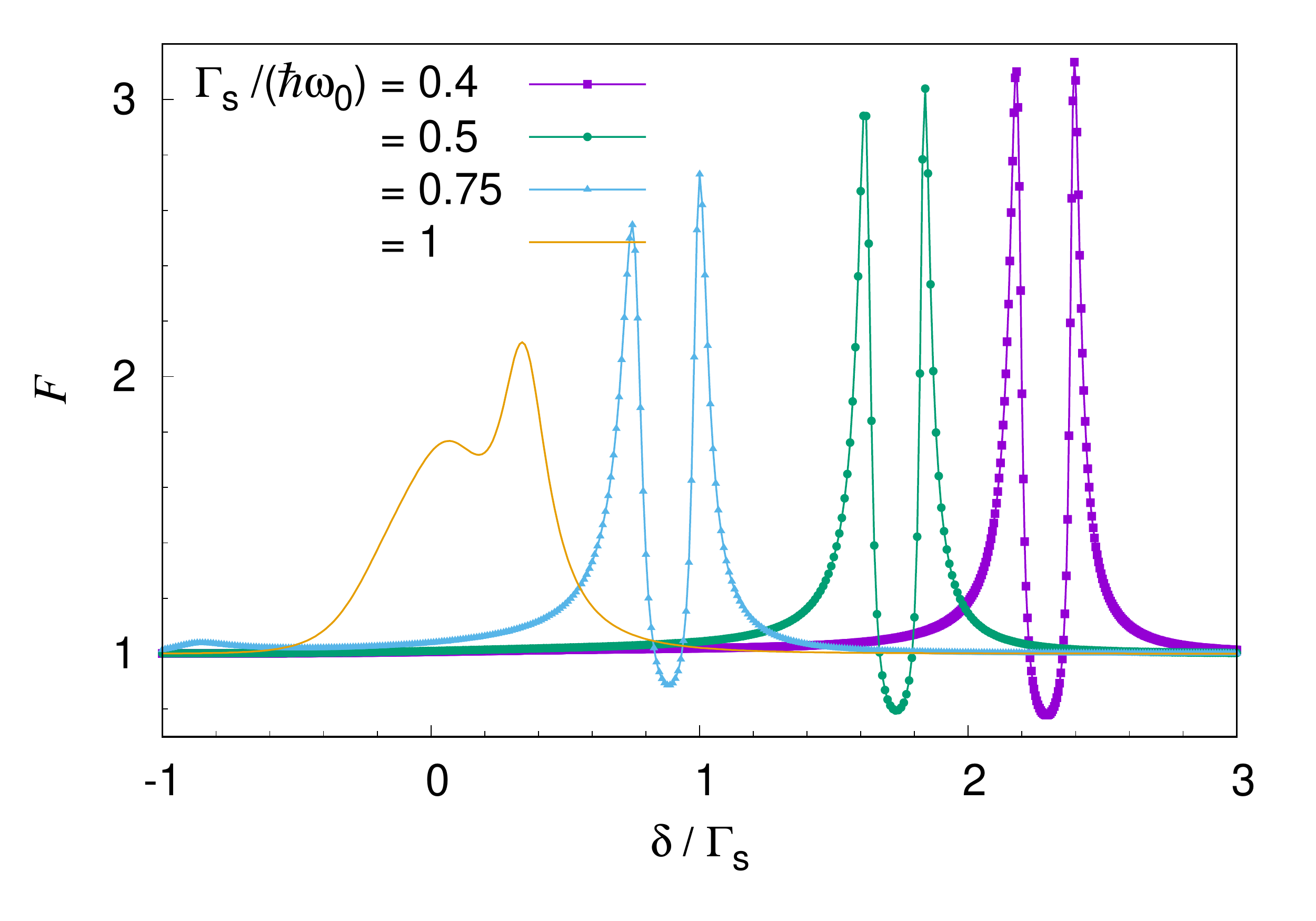}
\caption{
Fano factor of the oscillator mode obtained by solution of the master equation.
$F$ is plotted as  a function of the detuning $\delta=2\varepsilon+U$ in units of $\Gamma_S$ 
for different ratios of  $\, \Gamma_S/(\hbar\omega_0)$.
Parameters: $\lambda=0.01 \, \hbar \omega_0$ and $\kappa/\Gamma_N=0.02$.
}
\label{fig:DN}
\end{figure}
%
%
%
%

%
%
%
%
{\sl Results and discussion.}
%
One of the telltales of lasing is a large average occupation of the resonator mode. 
The average occupation of the oscillator mode as a function of the detuning is 
shown in Fig.~\ref{fig:noschdiss} together with the semiclassical 
result for small $\Gamma_N$, $\bar{n} =  [\Gamma_N/(2 \kappa)]\, \delta \, /{[ \delta^2 + \Gamma_S^2]}^{1/2}$.
As the semiclasical result corresponds to a situation where the parameters of the system are tuned to be always on resonance, 
it is an envelope to the curves corresponding to different values of $\Gamma_S$. 
Using the quantum master equation within the RWA, we find that $\bar{n}$ shows a peak as a function of the detuning at  
$\delta^2\approx (\hbar\omega_0)^2 -\Gamma_S^2$. 
The average occupation at the peak (on resonance) follows the semiclassical prediction, as can be seen in Fig ~\ref{fig:noschdiss}.

%
%
%
%
\begin{figure}[t!]
\includegraphics[scale=0.35,angle=270]{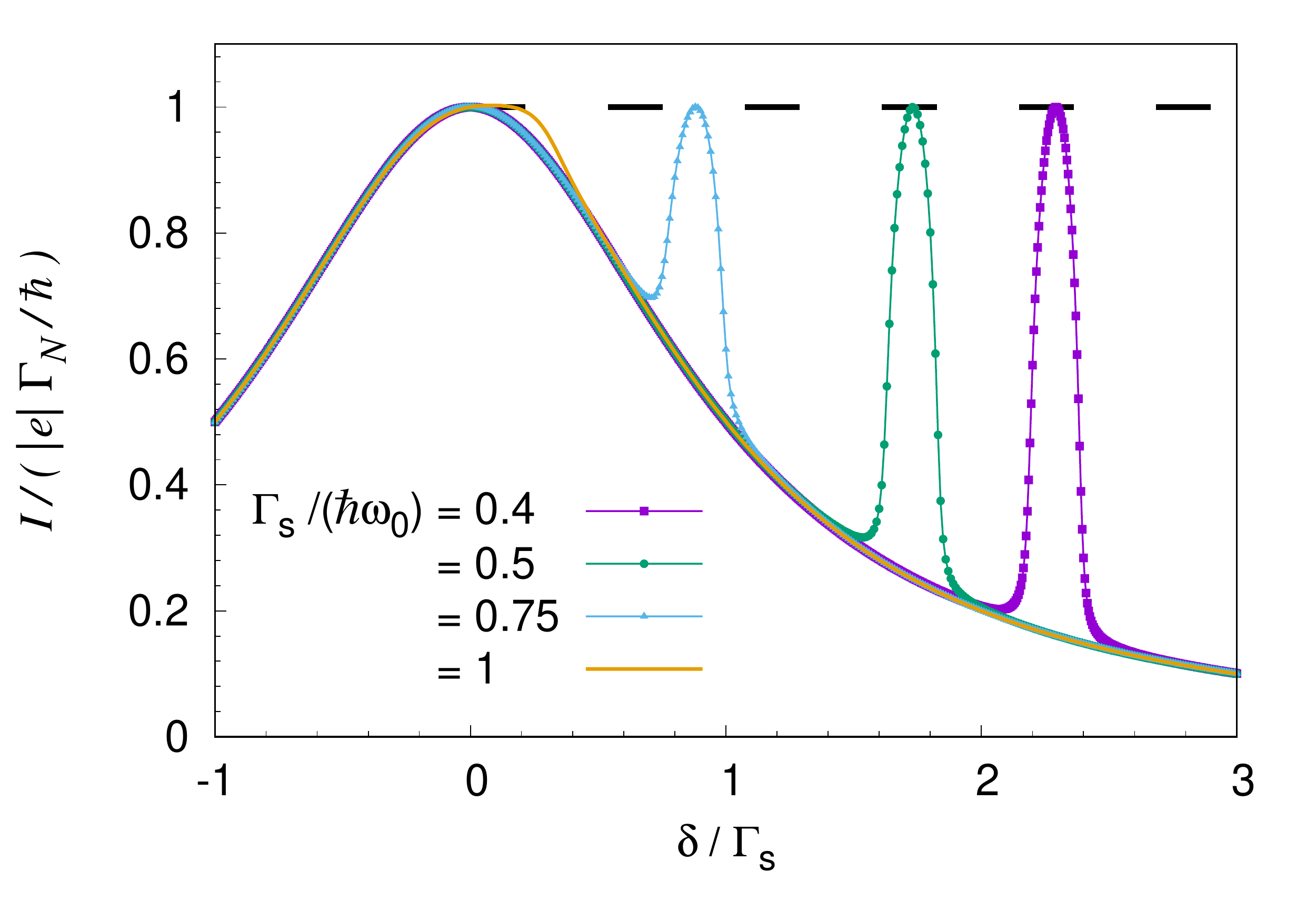}
\caption{
Current flowing through the quantum dot obtained by solution of the master equation.
$I$ is scaled with $|e| \Gamma_N/\hbar$ and is plotted as  a function of the detuning $\delta=2\varepsilon+U$ in units of $\Gamma_S$ 
for different ratios of $\Gamma_S/(\hbar\omega_0)$.
In the full lasing regime the current is pinned to the maximum value $|e| \Gamma_N/\hbar$ (dashed line).
Parameters: $\lambda=0.01 \, \hbar \omega_0$ and $\kappa/\Gamma_N=0.02$.
}
\label{fig:curr}
\end{figure}
%
%
%
%

In order to determine whether the peak in the average occupation of the resonator corresponds to a lasing state, 
we calculate the full distribution $p_n$. 
The Fano factor $F$ can be used to determine whether the resonator is in a lasing state. 
A value of 1 for $F$ indicates a Poissonian distribution while values smaller than 1 
correspond to a sub-Poissonian distribution. 
The Fano factor as a function of detuning is shown in Fig.~\ref{fig:DN}.
Interestingly 
$F$ computed by the master-equation approach drops below 1 in the vicinity of the peak clearly indicating sub-Poissonian lasing.
%
On the other hand, the Fano factor, obtained from the adiabatic approximation for coupling strength larger than the threshold, is approximately equal to 1 on resonance. 
Therefore we conclude that quantum fluctuations narrow the distribution $p_n$ on resonance, 
as it has already been found for different implementations of single atom lasers \cite{Mu:1992}.

Finally in Fig.~\ref{fig:curr}, we show the current as a function of the detuning. 
The current closely follows  the result for zero coupling to the oscillator except on resonance where it peaks
to its maximum value $|e|\Gamma_N/\hbar$ achievable in the large-gap regime.
The peak in the current indicates single atom lasing. 
The current behaviour can be explained following the arguments of Ref.~\cite{Braggio:2011}. 
In the absence of coupling to the resonator, for $\delta \gg  \Gamma_S$ the dot will be mostly stuck in the state $\ket{+}\approx\ket{D}$ and the bottle neck for transport is 
the cotunnelling of a Cooper pair to the superconductor with a rate $\propto \Gamma_S^2/\delta^2$ (refer also to Fig~\ref{fig:cartoon}(d)); this leads to a suppression of the current. 
The resonant coupling with the resonator removes this bottleneck by allowing the fast transition $\ket{+,n}\rightarrow\ket{-,n+1}\approx \ket{0,n+1}$. From the state $\ket{-,n+1}$ 
a  sequential tunnelling event with the normal lead  with rate $\propto \Gamma_N$, brings the dot in $\ket{\sigma,n+1}$ and a subsequent sequential tunnelling event to $\ket{+,n+1}$ 
(see Fig.~\ref{fig:cartoon}(e)).
In this case, the bottleneck for transport is the sequential tunnelling with the normal lead and the current reaches the same value as for $\delta=0$ and no coupling to the oscillator. 
We wish to emphasize that the process which restores the maximum value of the current is the very same that leads to lasing. 
Hence, the onset of lasing can be detected simply by a current measurement instead of measuring the population of the resonator.

%
%
%
%
{\sl Conclusions.}
%
In this letter we have shown that single-atom lasing can occur in a quantum dot strongly coupled to a superconductor.
Lasing is achievable within the reach of the experimental state of the art for these hybrid nanodevices.
Typical experimental values are $\lambda/(2\pi \hbar) \sim 50-100 \,\,\mbox{MHz}$ in single quantum dots \cite{Bruhat:2016br} and 
$\omega_0/(2\pi) \sim 7 \,\, \mbox{GHz} $ \cite{Bruhat:2016br,Stockklauser:2017bqa,Liu:2017}.
Large asymmetric tunnel coupling between the quantum dot and the superconductor has also been achieved $\Gamma_N \ll \Gamma_S$ \cite{Gramich:2015dk}.
$\Gamma_S$ can be tens of $\mu$eV reaching the typical microwave cavity frequency $\Gamma_S \lesssim \hbar \omega_0$  \cite{Gramich:2015dk}.
For computational reasons, we use a moderate damping coefficient for the intrinsic losses of the resonator.
%
Experimental devices with  large quality factor can have better performance, viz. $Q=\hbar\omega_0/\kappa$ 
with $Q\sim {10}^{4}$  \cite{Liu:2017}.
In this case, one can estimate lasing even for smaller values of the coupling constant \cite{footnote_large} than the one used here, $\lambda = 0.01\hbar \omega_0$.

\begin{acknowledgments}
G.R. is grateful for the warm hospitality at Victoria University of Wellington.
G.R. thanks  A.D. Armour for useful discussions. G.R. and M. G. thank K. G. Steenbergen for suggested improvements to the manuscript. 
G.R. acknowledges partial support from the German Excellence Initiative through the Zukunftskolleg 
and from the Deutsche Forschungsgemeinschaft (DFG) through the collaborative research center SFB 767.
\end{acknowledgments}

\bibliography{references.bib}

\begin{thebibliography}{39}
\bibitem [{com()}]{comparison}%
  \BibitemOpen
  \href@noop {} {}\bibinfo {note} {We have also performed a master equation
  approach beyond the RWA, with the full interaction Hamiltonian Eq.~(4). The
  results practically coincide with the RWA for the regime shown in this
  article (see Sup. Mat. \cite{sup_mat}).}\BibitemShut {Stop}%
\bibitem [{foo()}]{footnote_large}%
  \BibitemOpen
  \href@noop {} {}\bibinfo {note} {For instance, at resonance $2\epsilon_A
  =\hbar \omega_0$, and assuming $\Gamma_S \sim\hbar \omega_0$, see
  Fig.~\ref{fig:noschdiss},\ref{fig:DN},\ref{fig:curr} we have the scaling
  $\lambda_c/(\hbar \omega_0) \sim (1/\sqrt{Q}) {(
  \Gamma_N/\Gamma_S)}^{1/2}$.}\BibitemShut {Stop}%
\end{thebibliography}

\end{document}